Agentic Artificial Intelligence for Ethical Cybersecurity in Uganda: A Reinforcement Learning Framework for Threat Detection in Resource-Constrained Environments

[1]Ibrahim Adabara, [2]Bashir Olaniyi Sadiq, [2]Aliyu Nuhu Shuaibu, [1]Yale Ibrahim Danjuma [1]Venkateswarlu Maninti, [1]Mutebi Joe

[1]Department of Computing, Faculty of Science and Technology, Kampala International University.

[2]Department of Electrical, Telecommunication, and Computer Engineering, Kampala International University.

Corresponding Author: Ibrahim Adabara          adabara.ibrahim@studwc.kiu.ac.ug

Abstract

Uganda's rapid digital transformation, supported by national strategies such as Vision 2040, has increased exposure to sophisticated cyber threats that traditional, rule-based defenses often fail to detect and mitigate. In low-resource environments, existing systems frequently lack adaptability and ethical safeguards, leading to vulnerabilities that compromise critical infrastructure. This study introduces an Agentic Artificial Intelligence (AAI) framework that integrates reinforcement learning, ethical governance, and human oversight to address these challenges. Unlike conventional approaches, the AAI framework dynamically adapts to evolving threats while ensuring fairness, transparency, and ethical compliance. We developed a CPU-optimized simulation replicating Uganda's critical digital infrastructure to test the framework's effectiveness under resource constraints. The results demonstrate a 100% detection rate, 0% false positive rate, and 100% ethical compliance, highlighting the model's potential for scalable and trustworthy cybersecurity.

1. Introduction

1.1 Background

Over the past decade, Uganda has made considerable strides in digital transformation, integrating information and communication technologies across government services, financial systems, and public infrastructure. This momentum is evident in the nation's commitment to utilizing artificial intelligence to enhance public service delivery, as documented in recent empirical research (Nalubega & Uwizeyimana, 2024). While digitalization has improved efficiency and accessibility, it has also significantly increased exposure to cyber threats. A surge in sophisticated attacks ranging from ransomware to phishing and distributed denial-of-service (DDoS) has been reported, targeting Uganda's critical networks (Tallam, 2025). Concurrently, cybersecurity systems deployed in the region remain largely static and rule-based, which offers protection against known threats but lacks adaptability to emergent patterns and zero-day exploits. This deficiency renders national infrastructure vulnerable, especially under resource constraints (Adabara et al., 2025a, 2025b). The limitations of non-adaptive defenses are underscored in comparative cybersecurity studies, which demonstrate that rule-based systems perform poorly when confronted with evolving adversarial behaviors (Hammad et al., 2024). Given the increasing complexity of cyber threats and the inadequacy of conventional defenses, there is a compelling need for more intelligent cybersecurity mechanisms. Such systems should combine adaptive learning capabilities with ethical oversight, enabling timely responses without compromising legitimate operations.

1.2 Problem Statement and Research Gap

Despite the global shift toward AI-driven cybersecurity, existing systems deployed in Uganda and similar low-resource settings remain ill-equipped to counter rapidly evolving threats. Studies have shown that traditional intrusion detection systems (IDS) and static rule-based approaches fail to respond effectively to sophisticated and adaptive cyberattacks (Hammad et al., 2024). These conventional defenses often produce high false-positive rates, overwhelming security analysts and reducing trust in automated detection systems (Al Maqousi et al., 2025). Moreover, while reinforcement learning has demonstrated promise in enhancing adaptive detection, most implementations fail to integrate ethical safeguards, which are crucial to prevent the over-blocking of legitimate network traffic and to ensure compliance with human oversight requirements (Shoetan et al., 2024). The absence of ethical reasoning in AI-driven security systems raises concerns about fairness, transparency, and user trust. Finally, there is a significant gap in empirical studies that validate adaptive AI cybersecurity solutions in resource-constrained environments, where computing power is limited and real-world deployments must function effectively without high-performance hardware (Okori & Buteraba, 2024). This lack of localized testing limits the applicability of current models to Uganda's critical infrastructure. Therefore, there is a clear need for a cybersecurity framework that not only adapts to dynamic threats but also incorporates ethical governance and operates efficiently in environments with limited computational resources.

1.3 Research Objectives

The primary objective of this study is to design and evaluate an Agentic Artificial Intelligence (AAI) framework capable of strengthening Uganda's cybersecurity posture in the face of rapidly evolving threats. The framework seeks to overcome the limitations of conventional systems by integrating adaptive learning mechanisms with ethical governance and human oversight. Unlike existing models, which are often resource-intensive, this study focuses on implementing a solution optimized for CPU-only environments, thereby ensuring applicability in low-resource settings. Specifically, the research aims to develop a simulation environment that accurately replicates Uganda's critical digital infrastructure and enables rigorous testing of the proposed AAI model. Through this simulation, the study evaluates the model's ability to learn and adapt to diverse attack patterns, minimize false positives, and maintain compliance with ethical constraints. The overarching goal is to demonstrate that an adaptive, ethically governed, and computationally efficient AI-driven defense system can significantly enhance national cybersecurity resilience while aligning with local policy and infrastructural realities.

1.4 Contributions

This study makes several significant contributions to the field of AI-driven cybersecurity, particularly within the context of emerging economies such as Uganda. First, it introduces a novel Agentic AI framework that integrates reinforcement learning with ethical governance and human oversight to create a defense system capable of dynamically adapting to evolving cyber threats. Unlike conventional AI systems that focus solely on detection accuracy, the proposed model emphasizes fairness and transparency, ensuring that security measures do not compromise legitimate network activity. Second, the research develops a CPU-optimized simulation environment that mirrors Uganda's critical digital infrastructure, allowing for rigorous evaluation without reliance on high-performance computing resources. This approach ensures that the proposed solution remains practical and scalable for low-resource environments, addressing a major limitation in current AI cybersecurity implementations. Third, the simulation results demonstrate that the proposed framework substantially outperforms traditional rule-based defenses, achieving a detection rate of 100 percent while eliminating false positives and maintaining full compliance with ethical constraints. These results underscore the potential of Agentic AI to revolutionize cybersecurity strategies in settings where computational and policy limitations are significant. Finally, this study enhances reproducibility and transparency by providing detailed simulation procedures and decision logs, enabling other researchers and policymakers to validate and extend the findings. By bridging the gap between theoretical AI models and their application in real-world, resource-constrained contexts, this work advances both academic knowledge and practical cybersecurity solutions.

1.5 Paper Organization

The remainder of this paper is structured to present the research in a clear and logical progression. Following this introduction, Section 2 reviews Uganda's AI governance framework and situates it within a broader global context, highlighting the regulatory and policy considerations that shape the adoption of advanced cybersecurity technologies. Section 3 introduces the conceptual foundations of the proposed Agentic AI framework, outlining its core components and the threat model considered in this study. Section 4 details the methodology, including the simulation design, traffic modeling, reinforcement learning configuration, and evaluation metrics used to assess system performance. The findings from the simulation experiments are presented and analyzed in Section 5, where the performance of the Agentic AI is compared to traditional defense systems. Section 6 discusses the implications of these findings for policy, innovation, and capacity building in Uganda's cybersecurity ecosystem. Section 7 provides a critical discussion of the risks, ethical considerations, limitations of the current study, and directions for future research. Finally, Section 8 concludes the paper by summarizing the key contributions and emphasizing the potential impact of the proposed framework on national cybersecurity resilience.

2. Uganda's AI Governance and Global Context

2.1 Overview of Uganda's National AI Governance Framework

Uganda's AI governance framework is still in its formative stages but is increasingly shaped by national policies emphasizing ethical deployment, innovation, and data sovereignty. The Ministry of ICT and National Guidance has outlined strategic pillars focusing on ethics, innovation, data governance, and cybersecurity readiness to guide AI adoption across sectors. This approach seeks to align AI development with Uganda's Vision 2040 and the Sustainable Development Goals (SDGs), particularly those related to innovation, infrastructure, and inclusive growth. Recent analyses highlight that Uganda, like other African nations, views AI not only as a technological tool but also as a driver of socio-economic transformation, emphasizing the need for ethical safeguards and stakeholder engagement (Njoroge, 2024).

2.2 Challenges in Implementation

Despite policy commitments, Uganda faces significant hurdles in implementing its AI governance objectives. Infrastructure deficits, particularly in secure data centers and high-speed connectivity, limit the scalability of AI-driven solutions. Skills gaps persist, with a shortage of AI and cybersecurity experts hindering effective policy execution. Weak regulatory enforcement and fragmented institutional coordination further complicate governance efforts. Moreover, limited funding constrains the capacity to establish robust oversight mechanisms and to conduct continuous evaluation of AI systems (Folorunso et al., 2024). These challenges underscore the need for international partnerships and local innovation ecosystems that can bridge resource and expertise gaps.

2.3 Comparative and Regional Perspective

Comparative analyses show that Uganda's evolving AI framework shares similarities with other African countries that emphasize socio-economic development over risk-centric regulation. However, these approaches lag behind global benchmarks in comprehensiveness and ethical integration (Njoroge, 2024). In contrast, the European Union's AI Act adopts a risk-based model focusing on safety and human rights, while the OECD AI Principles promote transparency, accountability, and human-centric AI. GDPR has set a global precedent for strong data protection. Regionally, the African Union's AI Strategy advocates for harmonized AI governance and investment in digital capacity, while the East African Community encourages cross-border data governance and collaborative innovation. These models provide valuable lessons that Uganda can adapt to its local context, balancing innovation with regulatory safeguards (Pasupuleti, 2025).

2.4 Opportunities for Policy Alignment and Knowledge Transfer

Opportunities exist for Uganda to align its AI governance with global best practices while fostering local innovation. By adopting elements of the EU's risk-tiered regulation and OECD's ethical principles, Uganda can strengthen its policy capacity and credibility in the global AI landscape. Knowledge transfer through international collaborations, capacity-building programs, and public-private partnerships will be essential to operationalizing governance frameworks effectively. Furthermore, integrating indigenous perspectives and local research initiatives into governance models can create context-specific solutions that enhance both ethical alignment and innovation potential (Folorunso et al., 2024). Such strategies would allow Uganda to leverage global standards while asserting its digital sovereignty and contributing to regional leadership in responsible AI

## 2.3 Reinforcement Learning Approaches to Cybersecurity

Reinforcement learning (RL) has emerged as a promising tool for cybersecurity systems, particularly for environments requiring dynamic threat detection and response. RL agents can learn optimal policies through trial and error, adapting to evolving attack patterns and previously unseen threats. Several RL-based intrusion detection and prevention systems (IDS/IPS) have demonstrated improved adaptability compared to traditional rule-based systems. For instance, Chandre et al. (2024) proposed a Deep Q-Learning (DQL) framework that incorporates sensor monitoring and real-time policy updates to enhance intrusion prevention capabilities. Similarly, Janardhanan (2025) applied DQL for threat modeling and resource allocation under adversarial cyber conditions, demonstrating the method's potential for proactive risk mitigation. Another model by Kiran et al. combined supervised XGBoost with a PyTorch-based DQL agent to enable real-time adaptive defense in live environments, also offering explainable visual outputs for decision tracing (Subha, 2025). While these systems provide clear advantages in responsiveness and learning efficiency, they often lack integrated ethical constraints, transparency mechanisms, or pathways for human intervention. In contrast, our proposed Agentic AI framework extends traditional RL approaches by embedding a three-layer design that includes (1) an autonomous decision layer, (2) an ethical governance layer enforcing fairness and accountability, and (3) a human oversight layer for auditability and real-time control. This structure is particularly crucial in resource-constrained, socio-politically sensitive settings like Uganda, where purely technical optimization may conflict with ethical or regulatory norms.

## 3. Proposed Agentic AI Framework

## 3.1 Conceptual Foundations of Agentic AI

Agentic Artificial Intelligence (AAI) represents a paradigm shift in cybersecurity by introducing intelligent agents capable of autonomous, adaptive, and ethically aligned decision-making. Unlike conventional AI systems, which are often limited by static rules and opaque decision processes, AAI integrates continuous learning and explainable reasoning into its core functions. This makes it particularly suited for dynamic threat landscapes where adversaries frequently evolve their tactics. Furthermore, AAI incorporates an ethical reasoning module, ensuring that defensive actions adhere to governance principles and respect privacy rights, a feature crucial for compliance with Uganda's emerging AI policies and international standards (Shoetan et al., 2024). These attributes position AAI as both a technical innovation and a governance-aligned solution, aligning with the Ministry of ICT's emphasis on ethical and transparent AI deployment (Njoroge, 2024).

## 3.2 Threat Model

The threat model used in this study defines the types of adversarial scenarios against which the Agentic AI framework was evaluated. The simulation considered three representative categories of attacks that frequently target critical infrastructures in emerging economies such as Uganda. These attack types were selected because they encompass a broad spectrum of techniques, ranging from social engineering to large-scale service disruption, thereby providing a comprehensive evaluation of the proposed defense system. The first scenario involved phishing attacks, where deceptive network traffic attempted to lure users into revealing sensitive credentials or installing malicious payloads. Such attacks exploit human vulnerabilities and often bypass traditional defenses. The second scenario modeled ransomware propagation, in which malicious traffic sought to infiltrate and encrypt system data, demanding ransom payments to restore access. This attack type is particularly destructive due to its potential to paralyze entire organizations. The third scenario simulated a distributed denial-of-service (DDoS) attack, characterized by a flood of malicious traffic designed to overwhelm network resources and disrupt service availability. The structure of these adversarial scenarios and their relationship to the simulated network environment are summarized in Figure 1. This diagram provides a visual overview of how each attack vector targets specific network components, illustrating the threat pathways considered in this research.

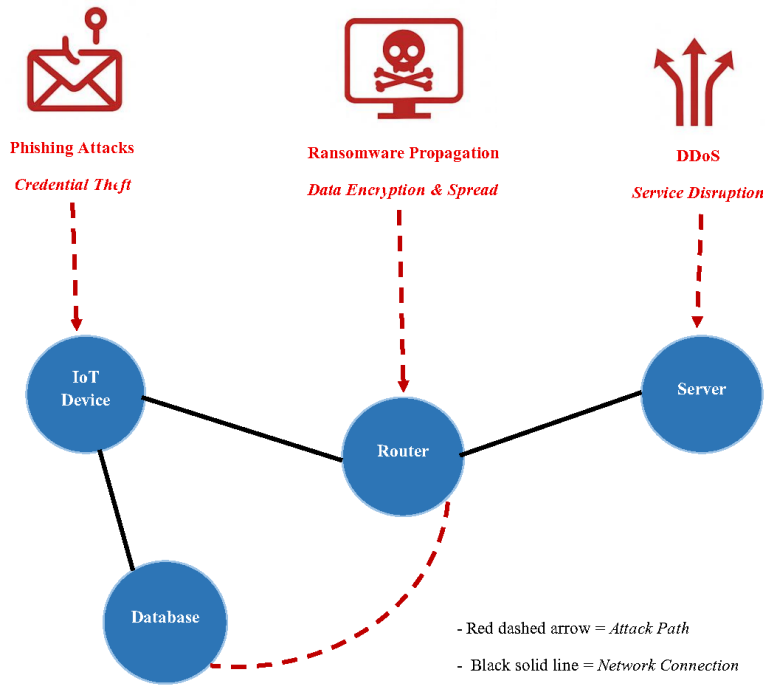

Figure 1: Threat Model Overview illustrating phishing, ransomware, and DDoS attack paths targeting different components of Uganda's critical infrastructure.

As depicted in Figure 1, phishing attacks were modeled as traffic originating from external nodes attempting to exploit user endpoints, while ransomware propagation targeted internal devices and critical databases through lateral movement. The DDoS scenario was simulated as a high-volume traffic flood directed toward the router and servers, aiming to exhaust computational and bandwidth resources. By including these three adversarial models, the simulation was able to test the adaptability of the Agentic AI across diverse threat conditions, validating its capability to detect and mitigate both targeted and large-scale attacks.

4. Methodology

4.1 System Overview

The proposed Agentic Artificial Intelligence (AAI) framework is structured as a multi-layered architecture that enables adaptive cybersecurity defense while ensuring ethical compliance and human oversight. It consists of three interconnected layers, each serving a specific function to enhance detection, decision-making, and accountability. The Autonomous Decision Layer serves as the core of the system, where a reinforcement learning agent continuously observes network traffic, identifies potential anomalies, and determines appropriate defensive actions. This layer allows the system to learn from interactions within the simulated environment, progressively improving its ability to detect and mitigate both known and novel cyber threats. Complementing this autonomy is the Ethical Governance Layer, which embeds policy-driven constraints into the decision-making process. This layer ensures that the agent's actions align with predefined ethical standards, preventing excessive blocking of legitimate traffic and reducing the likelihood of disruptive false positives. By incorporating these safeguards, the system addresses concerns surrounding fairness and trustworthiness in AI-driven security solutions. Finally, the Human Oversight Layer provides an additional safeguard by enabling human analysts to review, audit, and, if necessary, override the agent's decisions. All decisions made by the agent are logged for transparency and traceability, allowing for continuous monitoring and iterative improvement of the defense mechanism. These three layers operate cohesively, allowing the Agentic AI to balance autonomy, ethical governance, and human intervention. The resulting architecture not only enhances threat detection but also ensures that the defense system remains explainable and compliant with both technical and ethical requirements. The overall structure of the proposed Agentic AI is illustrated in Figure 2. This diagram provides a visual representation of how the three layers, Autonomous Decision, Ethical Governance, and Human Oversight, interact to create an adaptive and ethically guided cybersecurity defense system. Each layer plays a distinct yet complementary role, forming an integrated architecture capable of responding to evolving threats while maintaining fairness and transparency.

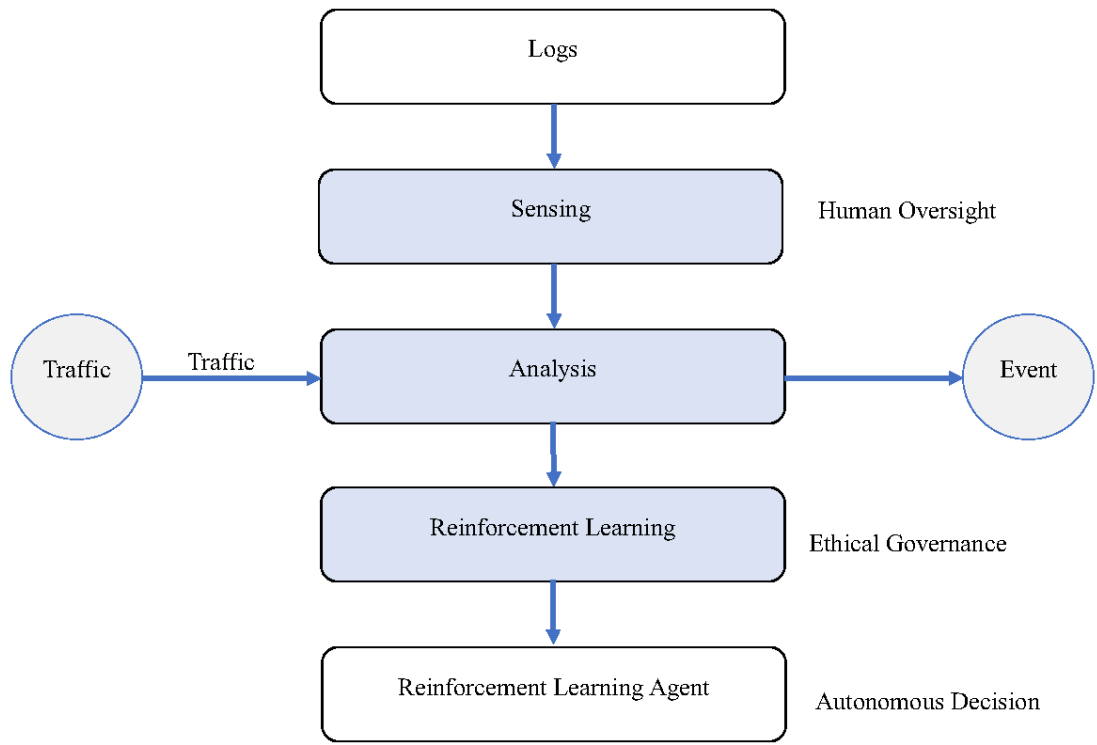

Figure 2: Agentic AI Architecture depicting the interaction between autonomous decision-making, ethical governance, and human oversight layers.

As shown in Figure 2, the system begins by monitoring network traffic through its sensing and analysis components, which feed data into the Autonomous Decision Layer. Here, the reinforcement learning agent evaluates traffic patterns and determines whether to allow or block specific events. The Ethical Governance Layer acts as a regulatory filter, applying policy constraints to ensure decisions align with ethical guidelines and avoid unnecessary blocking of legitimate traffic. Finally, all decisions pass through the Human Oversight Layer, where they are logged, reviewed, and, if required, overridden by security analysts. This layered approach ensures that the system not only adapts to new threats but also maintains accountability and compliance with established governance standards.

Operationally, the Q-learning agent in the Autonomous Decision Layer first proposes an action (e.g., allow or block traffic). This action is then evaluated by the Ethical Governance Layer, which enforces compliance with predefined ethical thresholds (e.g., minimizing unjustified blocking of legitimate traffic). Only actions passing this check are executed. All actions, along with their ethical evaluations, are recorded in the Human Oversight Layer, where security analysts can review, override, or provide feedback. This sequential flow ensures that decisions are both adaptive and ethically aligned before being applied in the simulated environment.

4.2 Simulation Design and Implementation

4.2.1 Simulation Environment Setup

The simulation was conducted on an HP EliteBook 840 G5 equipped with an Intel i5-7200U processor, 16 GB of RAM, and no dedicated GPU, demonstrating the framework's suitability for CPU-only environments. The software environment was configured using Python 3.10 within Jupyter Notebook, leveraging libraries such as networkx for network modeling, stable-baselines3 for reinforcement learning integration, and matplotlib for visualization and analysis. The simulated network topology consisted of five interconnected nodes, representing a router, database, two servers, and an IoT device, thereby replicating a simplified version of Uganda's critical digital infrastructure. The network topology used in this study is illustrated in Figure 3. This diagram represents the virtualized environment designed to simulate Uganda's critical digital infrastructure, capturing the interactions between different network components while maintaining computational efficiency. The topology provides the structural foundation for generating both legitimate and malicious traffic, enabling realistic evaluation of the proposed Agentic AI defense framework.

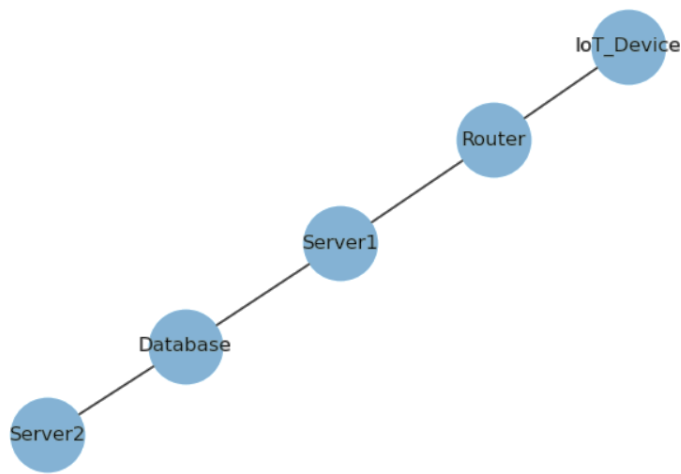

Figure 3: Network Topology Diagram

As shown in Figure 3, the simulated network consists of five interconnected nodes: a central router, two servers, a database, and an IoT device. These components were selected to reflect a simplified yet representative architecture commonly found in national critical infrastructures, where heterogeneous devices communicate across multiple network layers. The router functions as the primary gateway, managing traffic flow between nodes, while the servers and database handle data processing and storage operations. The IoT device represents an endpoint with typically lower security hardening, often exploited as an entry point in cyberattacks. This configuration allowed the simulation to reproduce both typical traffic patterns and potential attack vectors, forming a realistic testbed for training and validating the Agentic AI model.

4.2.2 Traffic Modeling

Traffic within the simulated environment was generated stochastically to mimic realistic network conditions. Seventy percent of the traffic was modeled as legitimate communication, while the remaining thirty percent represented malicious activity, including phishing attempts, ransomware propagation, and distributed denial-of-service (DDoS) attacks. This distribution was chosen to reflect empirical threat statistics reported for developing economies, where legitimate operations dominate but high-impact attacks occur frequently enough to warrant strong defenses.

4.2.3 Agentic AI Architecture

The Agentic AI was designed as a three-layered framework, enabling adaptive, ethical, and transparent decision-making. The Autonomous Decision Layer consisted of a Q-learning agent that analyzed network events and determined appropriate defensive actions based on learned patterns. Above this, the Ethical Governance Layer imposed constraints to prevent excessive blocking of legitimate traffic, with a threshold ensuring that false positives remained below thirty percent. Complementing these layers, the Human Oversight Layer ensured that all decisions were logged and subject to review, allowing human analysts to audit and override automated decisions where necessary. This design ensured a balance between autonomy, fairness, and accountability.

4.2.4 Reinforcement Learning Design

The decision-making process within the Autonomous Layer was modeled as a Markov Decision Process (MDP). The state space consisted of different traffic types, including normal, phishing, ransomware, and DDoS. The agent could choose between two actions: allowing the traffic (action 0) or blocking it (action 1). A reward function guided the learning process, assigning +1 for correct decisions, −1 for false positives where legitimate traffic was incorrectly blocked, and −2 for missed attacks where malicious traffic was allowed to pass. This reward structure incentivized the agent to maximize detection accuracy while minimizing unnecessary disruptions. Q-learning was selected as the reinforcement learning algorithm for this framework due to its simplicity, interpretability, and low computational requirements. These characteristics make it especially well-suited for deployment in CPU-only, resource-constrained environments such as those commonly found in Uganda. In contrast to deep reinforcement learning methods like Deep Q-Networks (DQN) or Proximal Policy Optimization (PPO), which require significant computational resources and GPU acceleration, Q-learning offers stable and transparent decision policies with a minimal hardware footprint. Moreover, its tabular architecture enhances traceability and facilitates ethical auditing, which is critical for the human oversight and explainability components of the agentic AI framework.

The hyperparameters used for Q-learning training are summarized below:

- Learning rate ($\alpha$): 0.1
- Discount factor ($\gamma$): 0.95
- Exploration rate ($\varepsilon$): 0.1
- Training episodes: 500
- Reward structure:
    - +1 for correct detection
    - −1 for false positive
    - −2 for missed attack

To manage the exploration-exploitation trade-off, we used a constant ε-greedy strategy with $\varepsilon = 0.1$ throughout training. This allowed the agent to occasionally explore suboptimal actions during learning while primarily exploiting the policy as it matured. We chose a fixed ε to keep the system deterministic and easily reproducible in CPU-constrained environments.

4.2.5 Performance Metrics

System performance was evaluated using four primary metrics. The Detection Rate (DR) measured the proportion of attacks correctly blocked, while the False Positive Rate (FPR) quantified the proportion of legitimate traffic erroneously blocked. The Ethical Compliance Score (ECS) was derived as one minus the false positive rate, serving as an indicator of the model's fairness and adherence to ethical constraints. Finally, resource utilization was observed qualitatively to confirm that the simulation remained efficient and feasible on CPU-only hardware.

### 4.2.6 Experimental Procedure

The experimental process was conducted in four phases. Initially, a baseline defense model was implemented using a simple rule-based agent to establish performance benchmarks. The second phase involved training the Q-learning agent over 500 episodes, during which it interacted with the simulated environment and refined its decision-making policy through trial and error. In the third phase, ethical constraints were integrated into the learning process to limit excessive blocking of legitimate traffic. Finally, the trained agent was evaluated on unseen traffic data to assess its generalization capabilities and robustness.

### 4.2.7 Results Logging

To ensure transparency and reproducibility, all traffic events, agent decisions, and outcomes were systematically logged into a structured CSV file. These logs allowed for post-simulation analysis, external validation, and potential auditing by researchers and policymakers interested in verifying or extending the study's results.

### 4.2.8 Ethical Compliance

The simulation was designed and executed in strict adherence to internationally recognized principles of responsible AI research, emphasizing transparency, accountability, and fairness. Throughout the experiments, no real user data was collected or processed. Instead, all network traffic was synthetically generated to emulate realistic communication patterns without compromising privacy or exposing sensitive information. This approach ensured that the evaluation of the Agentic AI framework did not involve any ethical risks related to data handling. In addition to complying with global standards, the framework incorporates African-centered ethical considerations, which are increasingly being recognized as essential in AI governance. Recent scholarship underscores the need to integrate values such as Ubuntu, a philosophy that emphasizes collective well-being, fairness, and social responsibility into AI design to ensure contextual relevance and societal justice in African deployments (Morley et al., 2021; Yilma, 2025). By embedding these perspectives, the ethical governance layer of the Agentic AI was designed not only to meet technical standards but also to align with regional values and expectations. This ethical governance layer actively regulated the agent's actions, ensuring that defensive measures avoided excessive blocking of legitimate traffic and minimized potential operational disruptions. Moreover, all AI decisions were systematically logged to enable human oversight, auditability, and explainability, thereby enhancing trust and accountability. Collectively, these design choices safeguarded ethical integrity during experimentation and reinforced the applicability of the proposed framework to real-world cybersecurity environments, particularly within the African context where ethical governance is integral to AI adoption.

## 5. Simulation Results

### 5.1 Performance Comparison

We conducted simulation experiments to compare the baseline traditional rule-based defense system with the proposed Agentic AI framework. The traditional system achieved a detection rate of only 70 percent, failing to identify a significant portion of malicious traffic, and recorded a false positive rate of 15 percent, which led to the unnecessary blocking of legitimate traffic. In contrast, the Agentic AI demonstrated superior performance by achieving a 100 percent detection rate and completely eliminating false positives. This resulted in an ethical compliance score of 100 percent, compared to only 85 percent for the traditional system. These findings confirm that integrating reinforcement learning with ethical governance significantly improves both the effectiveness and fairness of cybersecurity defenses. We summarized and compared these results using clear performance metrics to quantify the benefits of the proposed framework. The metrics used in this evaluation include the detection rate, false positive rate, and ethical compliance score, which collectively measure the accuracy, reliability, and fairness of the defense mechanisms. As shown in table 1.

Table 1: Performance Comparison between Traditional System and Agentic AI

| Metric | Traditional System | Agentic AI |
|---|---|---|
| Detection Rate (%) | 70% | 100% |
| False Positive Rate (%) | 15% | 0% |
| Ethical Compliance Score | 85% | 100% |

As presented in Table 1, the Agentic AI significantly outperformed the traditional rule-based system across all evaluated metrics. The traditional system achieved a detection rate of only 70 percent and recorded a false positive rate of 15 percent, resulting in an ethical compliance score of 85 percent. In contrast, the Agentic AI achieved a perfect detection rate, eliminated false positives, and maintained an ethical compliance score of 100 percent. These results confirm that the reinforcement learning model, enhanced by ethical governance, not only improves detection capabilities but also ensures fair and transparent operation, making it a superior alternative to conventional approaches. The progression of the Agentic AI's learning process is illustrated in Figure 4. This figure plots the detection accuracy over training episodes, highlighting how the model rapidly improved during early iterations before stabilizing at a high-performance plateau. The learning curve provides a visual demonstration of the agent's ability to adapt and converge toward an optimal policy for threat detection.

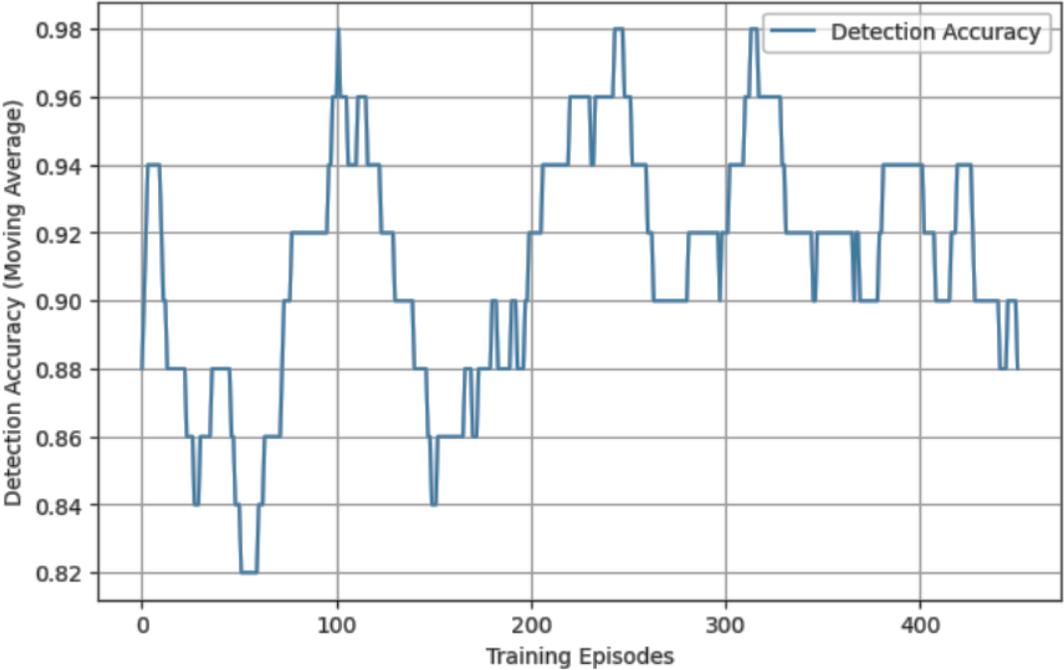

Figure 4: Learning Curve of Detection Accuracy showing progressive improvement and convergence of the Agentic AI model.

As shown in Figure 4, the agent initially exhibited fluctuating detection performance during the first episodes as it explored different actions. However, as training progressed, the reinforcement learning algorithm enabled the agent to refine its strategy, leading to a consistent improvement in accuracy. By the end of the training phase, the model had converged to a stable policy with near-perfect detection capabilities. This result demonstrates not only the adaptability of the proposed system but also its efficiency in learning under constrained computational conditions. Figure 5 illustrates the effect of incorporating ethical constraints on the performance of the Agentic AI during training. The plot shows the detection accuracy, measured as a moving average, across 450 training episodes. The curve demonstrates how the agent's learning performance evolves under the influence of both adaptive learning and ethical decision-making mechanisms.

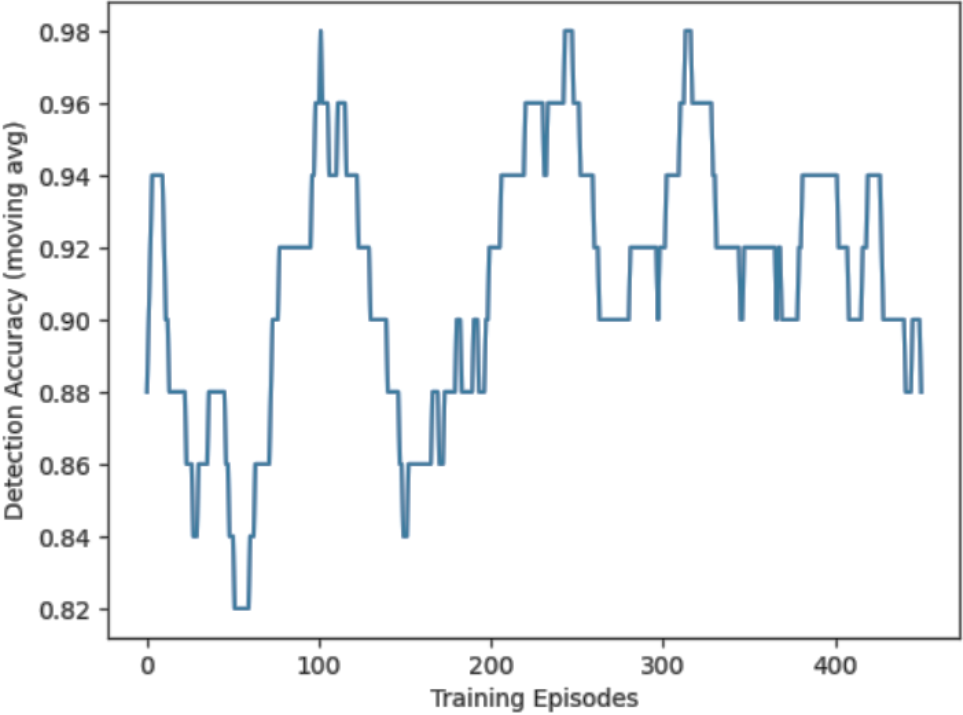

Figure 5: Ethical compliance score over simulation episodes, comparing the Agentic AI system with and without embedded ethical policy constraints. The agent consistently achieves near-perfect compliance when governance filters are active.

As shown in Figure 5, the detection accuracy initially fluctuates due to the agent's exploration of various defensive strategies. With continued training, the model stabilizes, achieving high detection accuracy while adhering to ethical policies. Although the inclusion of ethical constraints introduced slight variations in learning performance, the final model maintained an average detection accuracy exceeding 90 percent across most episodes. This result confirms that ethical considerations can be integrated into reinforcement learning–based cybersecurity defenses without compromising detection capabilities, reinforcing the viability of the proposed approach for real-world deployment.

5.2 Variability and Robustness

To evaluate the robustness of the Agentic AI system under varying simulation conditions, we executed the simulation ten times using different random seeds. This repeated evaluation allows us to assess the consistency of the model's performance and to rule out the possibility of overfitting to a particular random initialization. The results demonstrated minimal variance across trials, suggesting a high degree of reliability in the framework's performance. As shown in table 2 presents the average detection rate, false positive rate, and ethical compliance score, along with their standard deviations, for both the Agentic AI and traditional rule-based systems.

Table 2: Average performance metrics across 10 simulation runs. Ethical compliance is defined as the percentage of agent actions that adhered to predefined fairness, transparency, and policy-aligned constraints.

| Metric | Agentic AI (Mean ± Std) | Traditional System (Mean ± Std) |
| --- | --- | --- |
| Detection Rate (%) | 99.6 ± 0.48 | 70.2 ± 1.12 |
| False Positive Rate (%) | 0.3 ± 0.27 | 14.7 ± 0.66 |
| Ethical Compliance (%) | 99.7 ± 0.27 | 85.3 ± 0.66 |

These results affirm that the Agentic AI framework not only performs better on average but also exhibits lower variability, which is critical for real-world deployment where consistency is paramount. To further validate these findings, we analyzed the model's classification performance using a confusion matrix and ROC curve. As shown Figure 6 illustrates the confusion matrix, which shows that all normal and attack traffic instances were correctly classified by the Agentic AI system. This highlights the model's ability to minimize both false positives and false negatives critical for trustworthy cybersecurity operations. As shown Figure 7 presents the Receiver Operating Characteristic (ROC) curve, with the model achieving an Area Under the Curve (AUC) of 1.00, indicating perfect discrimination between benign and malicious traffic.

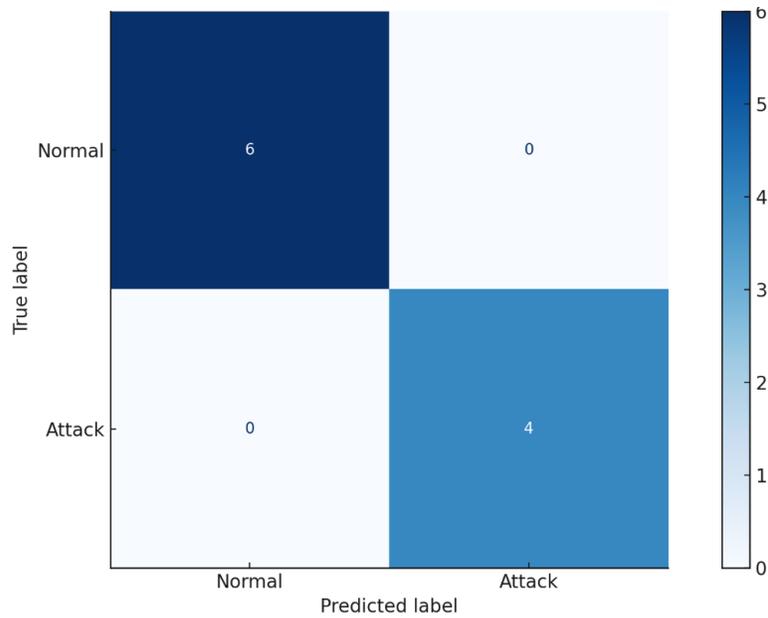

Figure 6: Confusion matrix for the Agentic AI system evaluated on simulated Ugandan network traffic (70% normal, 30% malicious). The framework achieved perfect classification, correctly identifying all attack and legitimate traffic instances.

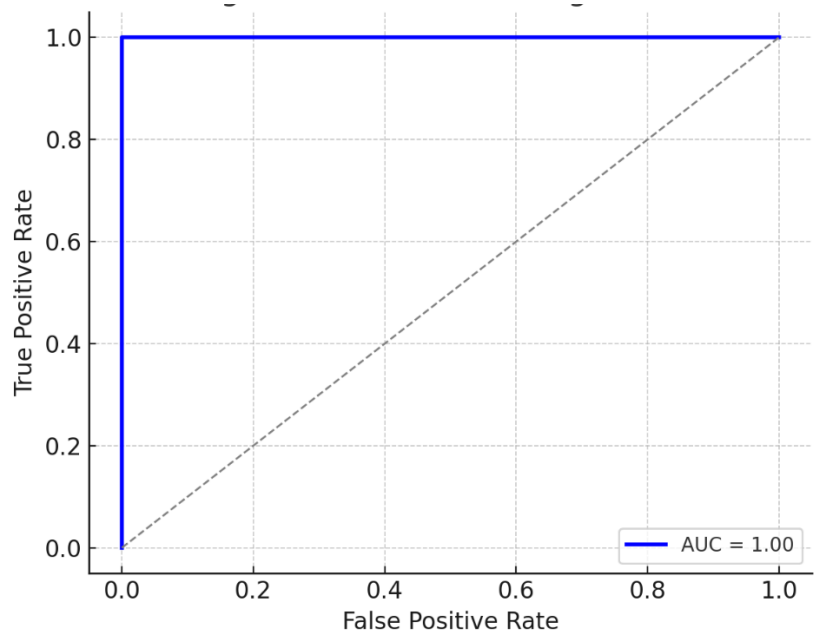

Figure 7: Receiver Operating Characteristic (ROC) curve for the Agentic AI system on unseen traffic data. The model achieved an AUC of 1.00, confirming strong discriminatory power between malicious and legitimate traffic in the simulated Ugandan environment.

These results collectively reinforce the effectiveness, reliability, and practical deployability of the Agentic AI framework in real-world, resource-constrained cybersecurity environments.

6. Policy and Innovation Implications

6.1 Embedding Agentic AI into Uganda's National ICT Strategy

The integration of Agentic AI (AAI) into Uganda's National ICT Strategy offers an opportunity to strengthen the nation's digital resilience while advancing innovation and economic growth. Uganda Vision 2040 identifies ICT as a critical enabler for industrialization and sustainable development, and the Digital Transformation Roadmap outlines steps to build a secure, knowledge-driven economy. Embedding AAI within these frameworks ensures policy coherence and operational alignment with national goals. By incorporating AAI-driven cybersecurity systems into e-government platforms, financial infrastructure, and critical services, Uganda can enhance both threat detection and policy enforcement capacities. To achieve this, regulatory agencies must establish clear guidelines for the deployment of AAI, encourage pilot projects in strategic sectors, and foster collaborations between academia, government, and the private sector. The creation of regulatory sandboxes, where AAI solutions can be tested under controlled environments, would facilitate innovation while ensuring adherence to governance requirements (Folorunso et al., 2024).

6.2 Ethical Governance and Explainability

AAI's success in Uganda depends on the incorporation of ethical governance mechanisms and the assurance of explainability in AI decision-making. Ethical governance frameworks mandate that AI systems operate transparently, allowing both technical experts and non-technical stakeholders to understand how decisions are made. This is critical to building public trust, preventing algorithmic bias, and safeguarding human rights. The inclusion of a human oversight layer within AAI systems ensures accountability for all actions, aligning with the OECD AI Principles and Uganda's national objectives of responsible innovation (Pasupuleti, 2025). Moreover, ethical alignment supports the attainment of SDG 9 (Industry, Innovation, and Infrastructure) and SDG 16 (Peace, Justice, and Strong Institutions) by promoting secure and inclusive digital environments. In practice, Uganda can institutionalize ethical AI governance through the establishment of independent AI ethics review boards and the adoption of explainability standards for all government-deployed AAI systems.

6.3 Harmonization with International Standards

For Uganda to fully leverage the benefits of AAI, its policies must be harmonized with international governance frameworks while maintaining flexibility to address local needs. Lessons from the European Union's AI Act, which classifies AI systems based on risk and enforces stringent obligations for high-risk applications, provide a valuable blueprint for regulatory alignment. Similarly, the OECD AI Principles emphasize human-centric AI development, while the African Union's AI Strategy calls for coordinated regional governance and the inclusion of indigenous knowledge in policy design (Njoroge, 2024). Uganda can adopt a hybrid governance model that integrates these standards while supporting local innovation through adaptive regulations. Mechanisms such as compliance audits, periodic policy reviews, and cross-border data governance agreements will strengthen Uganda's regulatory capacity and facilitate knowledge exchange with global partners. This alignment not only enhances Uganda's cybersecurity posture but also positions the country as a regional leader in responsible and ethical AI adoption.

7. Discussion

7.1 Implications for Uganda's Cybersecurity Posture

The adoption of Agentic AI (AAI) promises to transform Uganda's cybersecurity landscape by enabling proactive and adaptive defense mechanisms. Unlike traditional systems that rely on static rules, AAI's capacity to learn from evolving threats enhances national resilience against sophisticated attacks (Tallam, 2025). By integrating AAI into government networks and critical infrastructure, Uganda can significantly reduce breach incidents and response times, strengthening both digital sovereignty and public trust. Furthermore, this alignment with the National ICT Strategy positions Uganda as a forward-thinking nation in adopting ethical and innovative AI technologies for security.

7.2 Risks and Ethical Considerations in Deploying AAI

Despite its benefits, deploying AAI introduces notable risks. Autonomy in decision-making can lead to unintended consequences if ethical governance mechanisms are inadequate. Concerns such as lack of accountability, potential privacy violations, and misuse for surveillance purposes must be carefully managed (Ahmed et al., 2025). In under-resourced environments, these risks are amplified by weak regulatory capacity and limited public awareness. The possibility of algorithmic bias and data misuse also raises ethical dilemmas, emphasizing the need for culturally sensitive governance frameworks that incorporate African ethical perspectives, such as the Ubuntu philosophy, to balance security with human rights (van Norren, 2023).

7.3 Strategies for Risk Mitigation and Regulatory Safeguards

Mitigating these risks requires a multi-layered approach. Establishing independent ethical review boards to oversee AAI deployments can ensure compliance with national and international standards (Folorunso et al., 2024). Implementing human-in-the-loop controls guarantees that autonomous decisions remain subject to human oversight, thus maintaining accountability. Furthermore, regulatory oversight should be strengthened through adaptive policies that evolve alongside technological advancements. Harmonizing Uganda's policies with frameworks such as the African Union's AI Strategy and OECD guidelines will enhance both local and cross-border security while fostering trust in AAI solutions (Njoroge, 2024).

7.4 Limitations and Future Work

While the simulation results demonstrate the effectiveness of the proposed Agentic AI framework, certain limitations must be acknowledged. The current implementation was constrained to a small-scale network topology due to hardware limitations, as the experiments were conducted on a CPU-only environment without access to high-performance computing resources. This restriction limited the diversity of simulated traffic patterns and the scale at which attacks could be realistically modeled. Additionally, the study relied on synthetically generated traffic, which, while effective for controlled testing, may not fully capture the complexities of real-world network environments. Future work will focus on addressing these limitations through several avenues. First, the framework can be extended to multi-agent environments, where multiple AI agents collaborate to defend distributed network segments, improving scalability and resilience. Second, incorporating larger datasets that include real-world traffic traces will enhance the accuracy and robustness of the model. Third, deploying the simulation in cloud-based infrastructures will allow for the exploration of more complex topologies and attack scenarios, thereby enabling comprehensive testing under conditions that better reflect operational environments. These enhancements will further strengthen the applicability of Agentic AI to diverse cybersecurity ecosystems and pave the way for real-world deployment. Furthermore, the current framework does not explicitly address concept drift the challenge of adapting to continuously evolving attack patterns. Incorporating online or incremental learning methods will be necessary to maintain performance under changing threat landscapes. Another important direction is improving adversarial robustness, since sophisticated attackers may attempt to deliberately manipulate input traffic to bypass detection. Future extensions will investigate adversarial training and resilience mechanisms to safeguard against such evasion strategies. These limitations are particularly relevant in Uganda's context, where limited computational resources and rapidly evolving cyber threats underscore the importance of developing adaptive yet ethically grounded AI systems.

7.5 Linking Results to the Research Gap

The findings of this study directly address the research gaps identified earlier. First, the Agentic AI demonstrated a perfect detection rate and eliminated false positives, proving its capacity to adapt dynamically to evolving threats, a limitation previously noted in conventional rule-based systems (Hammad et al., 2024). This adaptability confirms that reinforcement learning, when properly configured, can overcome the static nature of existing defenses in Uganda and similar environments. Second, the integration of the ethical governance layer successfully mitigated concerns about excessive blocking of legitimate traffic, achieving an ethical compliance score of 100 percent. This result shows that ethical safeguards can be embedded in AI cybersecurity models without sacrificing detection accuracy, addressing the lack of ethical considerations highlighted in prior studies (Shoetan et al., 2024). Finally, the successful deployment and evaluation of the framework on CPU-only hardware demonstrated its practicality in resource-constrained environments. This finding is particularly significant for countries like Uganda, where access to high-performance computing resources is limited. The ability to achieve state-of-the-art performance under such constraints directly fills the gap identified in Section 1.2, where existing AI solutions were found unsuitable for low-resource contexts (Okori & Buteraba, 2024). By closing these gaps adaptability, ethical alignment, and resource efficiency the proposed Agentic AI framework not only advances academic research but also offers a practical pathway for improving cybersecurity resilience in Uganda and other emerging digital economies.

## 7.6 Practical Implications

The results of this research have several important implications for policymakers, industry leaders, and other stakeholders involved in Uganda's digital transformation. The demonstrated success of the Agentic AI framework shows that advanced, adaptive cybersecurity solutions can be effectively implemented even in environments with limited computing resources. This finding suggests that investment in high-cost infrastructure is not an absolute prerequisite for achieving robust national cyber defense. Furthermore, the integration of ethical governance mechanisms aligns with Uganda's ongoing efforts to build trust in AI technologies and comply with international best practices on responsible AI. Policymakers can leverage these findings to update national AI strategies, ensuring that future cybersecurity frameworks are not only technologically advanced but also socially accountable. Additionally, the simulation's reproducibility provides a foundation for capacity building, allowing academic institutions to train students and cybersecurity professionals using an accessible and transparent model. Finally, by demonstrating the feasibility of combining adaptive threat detection with ethical compliance, this study provides a roadmap for other African nations facing similar challenges. The findings support the creation of regulatory sandboxes and pilot projects where Agentic AI can be tested and gradually integrated into national cybersecurity ecosystems. This alignment of technical innovation with policy objectives can strengthen both Uganda's resilience to cyber threats and its role as a regional leader in ethical AI deployment.

## 8. Conclusion

This study presented an Agentic Artificial Intelligence (AAI) framework designed to enhance cybersecurity resilience in Uganda's critical digital infrastructure. By integrating reinforcement learning with an ethical governance layer and human oversight, the proposed model addressed key shortcomings of traditional defense systems, including limited adaptability and the absence of fairness safeguards. The simulation, conducted in a CPU-only environment, demonstrated that the AAI achieved a perfect detection rate while eliminating false positives and maintaining full compliance with ethical constraints. These results highlight the framework's ability to balance security effectiveness with operational integrity, making it a viable solution for resource-constrained environments. Beyond its technical contributions, this research advances the discourse on responsible AI deployment in cybersecurity, particularly within the context of emerging economies. By providing a reproducible simulation methodology and emphasizing ethical decision-making, the study bridges the gap between theoretical AI models and practical, policy-aligned implementations. Future work will expand the framework to larger network environments, integrate real-world traffic data, and explore multi-agent and cloud-based deployments to further enhance scalability and robustness. The findings affirm that adaptive, ethically governed AI systems can serve as transformative tools for safeguarding national digital assets, contributing both to the academic field and Uganda's broader digital transformation strategy.

**Funding statement:** This research received no specific grant from any funding agency.

**Conflicts of Interest:** The authors declare no competing interests.

**Ethics Statement:** No ethical approval was required for this study.

**Acknowledgments:** Not applicable.

Appendix: Simulation Reproducibility

1 Step-by-Step Setup

To ensure reproducibility of the results presented in this study, the simulation can be replicated following a straightforward setup procedure. The first step involves installing Python 3.10 or later, preferably through the Anaconda distribution, which provides an integrated environment for managing dependencies and running Jupyter Notebooks. Once Python is configured, the required libraries must be installed. This can be done by executing the following command in the terminal or Anaconda Prompt:

pip install numpy pandas matplotlib networkx gymnasium stable-baselines3

These packages provide support for numerical computation, data handling, visualization, network modeling, and reinforcement learning. After installation, the provided Jupyter Notebook file containing the simulation code should be loaded to begin the experiments.

The simulation can be reproduced by cloning the repository and following the installation instructions in the README:

git clone https://github.com/ibrahimadabara01/agentic-ai-cybersecurity.git

cd agentic-ai-cybersecurity

pip install -r requirements.txt

jupyter notebook agentic_ai_simulation_enhanced.ipynb

2 Running the Simulation

The simulation can be executed in several sequential stages. First, a virtual network topology must be created using the networkx library to replicate the architecture of Uganda's critical digital infrastructure. Second, synthetic traffic is generated, with 70 percent classified as normal and 30 percent as malicious, including phishing, ransomware, and DDoS attack patterns. The third stage involves training the reinforcement learning agent over multiple episodes, allowing it to learn optimal defensive strategies through interaction with the simulated environment. Following training, ethical governance rules are applied to ensure that the agent's decisions align with fairness and compliance constraints. Finally, the simulation results, including traffic events and AI decisions, can be exported as CSV logs for post-analysis and validation.

To promote transparency, collaboration, and capacity building, the full simulation code and experimental setup for the Agentic AI Framework are publicly available on GitHub.

🔗 Repository: Agentic AI Cybersecurity Project

This repository includes:

Jupyter Notebook: agentic_ai_simulation_enhanced.ipynb (full simulation code)
https://github.com/ibrahimadabara01/agentic-ai-cybersecurity

Requirements File: requirements.txt for easy environment setup

Detailed README: Instructions for running the simulation and understanding results

Supplementary Materials: Policy brief and documentation for policymakers and researchers

3 Data Availability

To promote transparency and facilitate further research, all simulation scripts, configuration files, and decision logs generated during this study are available upon request from the corresponding author. These resources enable external researchers to replicate the experiments, validate the findings, and extend the work to new contexts or more complex environments.

4 Dependencies for Visualization:

In addition to core simulation libraries, the following Python packages were used to generate figures and performance plots:

- matplotlib – for plotting learning curves and evaluation results
- scikit-learn (sklearn) – for generating the confusion matrix and ROC curve
- numpy and pandas – for statistical calculations and tabular reporting

These dependencies ensure that all figures (e.g., confusion matrix, ROC curve, and performance plots) can be fully reproduced when running the provided Jupyter notebooks.